\documentclass[aps,prl,twocolumn,showpacs,preprintnumbers,nofootinbib,amsmath,amssymb,superscriptaddress,nobalancelastpage]{revtex4}

\usepackage{graphicx}
\usepackage{dcolumn}
\usepackage{bm}

\newcommand{\beq}{\begin{equation}}
\newcommand{\eeq}{\end{equation}}

\newcommand{\Neq}{N_{\rm eq}}

\newcommand{\Msun}{M_\odot}

\newcommand{\mchi}{m_{\chi}}

\newcommand{\rhosun}{\rho_\odot}

\newcommand{\Cc}{C_\mathrm{c}}
\newcommand{\Ca}{C_\mathrm{a}}

\def\yr {\ifmmode \,\, {\rm yr} \else yr \fi}

\def\nuID{\nu {\rm ID }}

\begin{document}

\title{Galactic Substructure and Energetic Neutrinos from
the Sun and the Earth}

\author{Savvas M. Koushiappas}
\affiliation{Department of Physics, Brown University, 182 Hope
Street, Providence, RI 02912} \email{koushiappas@brown.edu}
\author{Marc Kamionkowski}
\affiliation{California Institute of Technology, Mail Code
350-17, Pasadena, CA 91125}\email{kamion@tapir.caltech.edu}

\begin{abstract}
We consider the effects of Galactic substructure on energetic
neutrinos from annihilation of weakly-interacting massive
particles (WIMPs) that have been captured by the
Sun and Earth. Substructure gives rise to a time-varying capture rate and
thus to time variation in the annihilation rate and resulting
energetic-neutrino flux.  However, there may be a time lag
between the capture and annihilation rates.  The
energetic-neutrino flux may then be determined by the density of
dark matter in the Solar System's past trajectory,
rather than the local density.  The signature of such an effect
may be sought in the ratio of the direct- to indirect-detection
rates.
\end{abstract}

\maketitle


Numerous experimental probes have confirmed indirectly the
presence of a yet unknown form of gravitationally-interacting
matter in Galactic halos that contributes roughly 20\% of the
total cosmic energy density.  It is generally assumed that
``dark matter'' is in the form of some yet undiscovered
elementary particle. Among the plethora of proposed theoretical
particle dark-matter candidates, weakly interacting massive
particles (WIMPs) are favored because they provide, quite
generally, the correct relic abundance and because they may be 
experimentally accessible in the near future. WIMPs arise
naturally in supersymmetric extensions (SUSY) of the Standard
Model \cite{Jungman:1995df} as well as in models with Universal
Extra Dimensions (UEDs) \cite{Hooper:2007qk}.

The two principle avenues toward dark-matter detection are
direct detection (DD) via observation of the recoil of a
nucleus, when struck by a halo WIMP, in a low-background
experiment \cite{Goodman:1984dc,Griest:1988ma}; and neutrino
indirect detection ($\nuID$) via observation of energetic
neutrinos from annihilation of WIMPs that have been captured in
the Sun (and/or Earth)
\cite{neutrinos,Gould:1987ir}.

The DD rate is proportional to the local dark-matter density.
The $\nuID$ rate is proportional to the rate at which WIMPs
annihilate in the Sun, which in turn depends on an
integral of the square of the dark-matter density over the
volume of the Sun. However, the WIMPs depleted in the Sun by
annihilation are replenished by the capture of new WIMPs.  In
most cases where the $\nuID$ signal is large enough to be
detectable, the
timescale for equilibration of capture and annihilation is small
compared with the age of the Solar System. The
$\nuID$ rate is then also determined by the local dark-matter
density.  Since the capture rate is controlled by the same
elastic-scattering process that occurs in DD, the DD and $\nuID$
rates are roughly proportional \cite{Kamionkowski:1994dp}.

In this Letter we investigate the effects of
Galactic substructure on this canonical scenario.  
Analytic arguments and numerical simulations suggest
that realistic Galactic halos should have significant
substructure, remnants of smaller halos produced in early stages
of the structure-formation hierarchy (which may themselves house
remnants of even smaller structures, and so on)
\cite{substructure}.  
Theoretical arguments suggest that the substructure may
be scale invariant \cite{Kamionkowski:2008vw} with subhalos
extending all the way down to
sub-Earth-mass scales \cite{earthmass}. The local
dark-matter density of different locations at similar
Galactocentric radii in the Milky Way may thus differ by a few orders
of magnitude.  The analytic descriptions of substructure are
rough, and the simulations are limited by finite resolution, and
this motivates the pursuit of avenues toward empirically probing
the existence of substructure.  

The purpose of this Letter to show that measurements of the
ratio of DD to $\nuID$ rates can be used to test for Galactic
substructure. If there is Galactic substructure, then the
dark-matter density at the position of the Solar System may vary
with time.  There is a finite time lag between capture and
annihilation, and so
the current energetic-neutrino flux may be determined not by the
local dark-matter density, but rather by the density of
dark matter along the past trajectory of the Solar
System.  The ratio for the $\nuID$/DD rate may thus differ from
the canonical prediction.  A departure from the canonical ratio
would thus, if observed, provide information about Galactic
substructure.  Since the equilibration timescale in the Earth is
generally different from that in the Sun, additional information
might be provided by observation of energetic neutrinos
from WIMP annihilation in the Earth.

To illustrate, we suppose the WIMP has a scalar
coupling to nuclei, but the formalism can be easily generalized
to spin-dependent WIMPs.  Then the DD rate for a WIMP of mass
$m_\chi$ from a target nucleus of mass $m_i$ is
\cite{Jungman:1995df,Kamionkowski:1994dp},
\begin{eqnarray}
\label{eqn:direct}
     R_{\rm DD}^{\rm sc} &=& 2.2\times10^5 \, {\rm
     kg}^{-1}\,{\rm yr}^{-1} \, 
     \rho_{\chi,0.3}  \eta_c(m_\chi,m_i)
     \left(\frac{\mchi}{100\,{\rm GeV}}\right) 
       \nonumber \\
     &\times& 
      \left( \frac{m_i}{100\,{\rm GeV}} \right) 
     \left( \frac{m_i}{\mchi + m_i} \right)^2 \sigma_{40},
\end{eqnarray}
where $\rho_{\chi,0.3}$ is the local dark-matter density in
units of 0.3~GeV~cm$^{-3}$, and $\eta_c(m_\chi,m_i) $ (given in
Ref.~\cite{Griest:1988ma}) accounts for form-factor
suppression. Here, $\sigma_{40}$ is the cross section for
WIMP-nucleon scattering in units of $10^{-40}$~cm$^2$.

The flux of upward muons induced in a neutrino telescope by
neutrinos from WIMP annihilation in the Sun is
\begin{eqnarray}
\label{eqn:neutrinos}
     \Gamma_{\nu,0} &=& 7.3\times10^5\, {\rm km}^{-2}\, {\rm
     yr}^{-1}\, (N/\Neq)^2 \rho_{\chi,0.3} \nonumber \\ 
      &\times& f(\mchi) [\xi(\mchi)/0.1]
 (\mchi/100\,{\rm GeV})^2 \sigma_{40},
\end{eqnarray}
while the corresponding flux from the Earth is obtained by
replacing the prefactor of Eq.~(\ref{eqn:neutrinos}) by 
$15\, {\rm km}^{-2}\, { \rm yr}^{-1}$. 
The function $f(\mchi)$ varies over the range $5 \le f(\mchi)
\le 0.5$ over the mass range $ 10 \le \mchi/{\rm GeV} \le 1000$
for the Sun (with a slightly larger range for the Earth), while
the function $\xi(\mchi)$ is in the range $\sim 0.01-0.3$ over
the same mass range.

The factor $N/N_{\rm eq}$ in Eq.~(\ref{eqn:neutrinos}) quantifies
the number of WIMPs in the Sun.  Once
WIMPs are captured in the Sun, they
accumulate deep within the solar core,
where they may annihilate to a variety of heavy Standard Model
particles which then decay to produce high-energy neutrinos
(which may escape the Sun).  The time $t$ evolution of the
number $N$ of WIMPs in the Sun, is governed by the differential
equation,
\begin{equation}
\label{eqn:riccati} 
     dN/dt = \Cc - \Ca N^2,
\end{equation}
where $\Cc$ is the capture rate of WIMPs by the Sun, and $\Ca
N^2$ is twice (because each annihilation destroys two WIMPs) the
effective annihilation rate. If both $\Cc$ and $\Ca$ are
constant and the initial condition is $N(t=0) \equiv N_0$, the
solution to this equation is
\begin{equation} 
\label{eq:Nt1}
     N(t) = \sqrt{ \frac{ \Cc }{\Ca}} \frac{ e^{t/\tau} - \gamma
     e^{-t/\tau}}{e^{t/\tau}+\gamma e^{-t/\tau}}, 
     \end{equation}
where
\begin{equation} 
     \gamma \equiv \frac{ 1 - N_0 \sqrt{\Ca / \Cc}}{1 + N_0 \sqrt{\Ca / \Cc}} \le 1, 
\end{equation}
and $\tau = 1 / \sqrt{\Cc \Ca}$ is the equilibration timecale. 
After a time $t\gtrsim \tau$, the number $N$ approaches
$     \Neq \equiv N(t \gg \tau) = \sqrt{ \Cc / \Ca}$,
and the annihilation rate $\Gamma_\nu$ becomes equal to (one half) the
capture rate, $\Gamma_\nu = \Ca N^2 / 2 = \Cc / 2$.

\begin{figure}[t]
\resizebox{!}{7cm}{\includegraphics{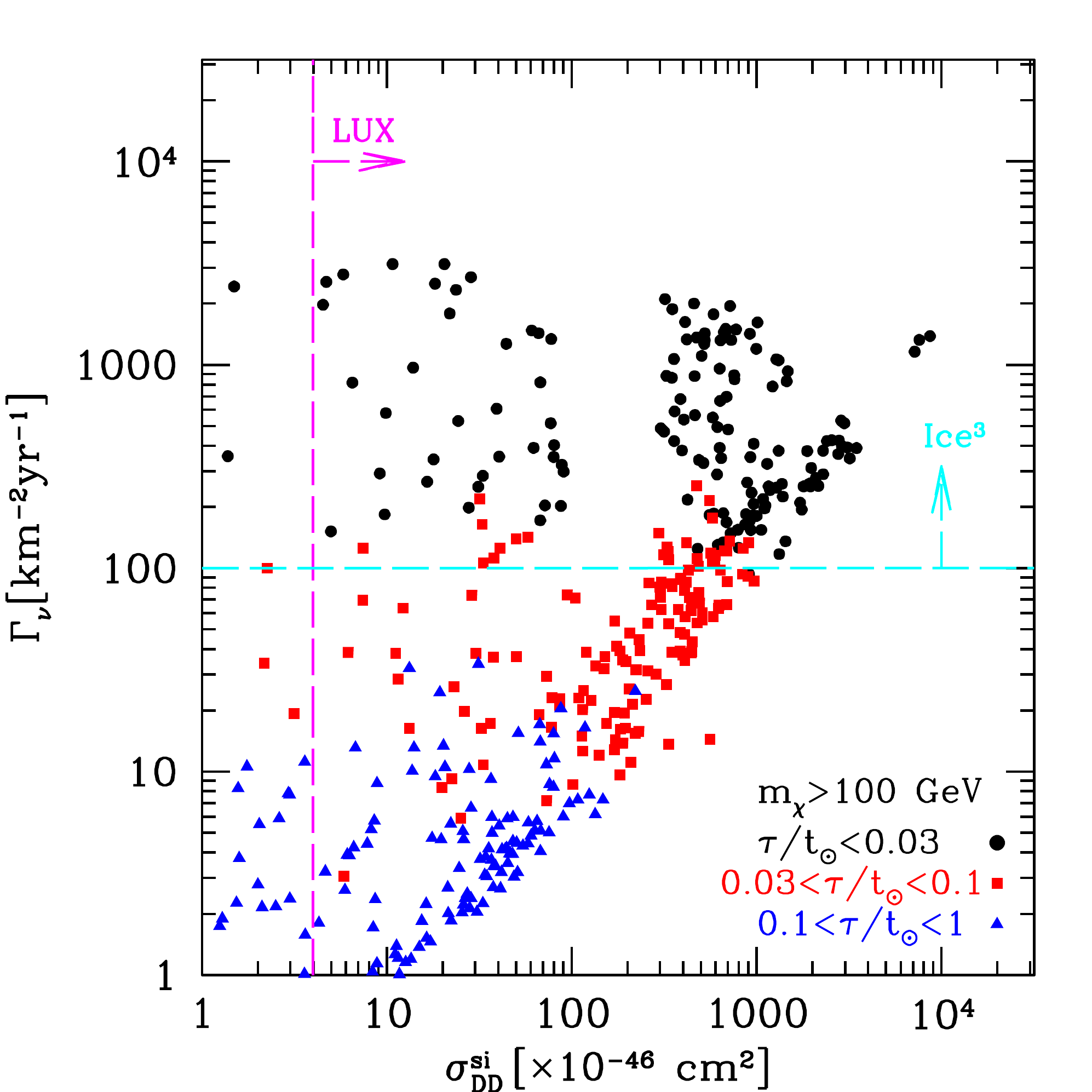}} 
\caption {\small The flux of energetic neutrinos from the Sun
     versus the rate for direct detection.
     Each point denotes a supersymmetric model with the correct
     relic density and consistent with experiment.
     The different symbols indicate the timescale for
     equilibration between capture and annihilation in the Sun.
     The horizontal line indicates a flux-threshold target for
     future $\nuID$ experiments (IceCube+DeepCore
     \protect\cite{DeYoung:2009er} and
     the vertical line a rate-threshold target for DD in a 3-ton
     liquid-xenon detector \cite{lux}. For current bounds see 
     \cite{Ahmed:2008eu}.}
\label{fig:PP}
\end{figure}

The  capture rate $C_c$ and annihilation coefficient $C_a$, and
thus the equilibration timescale $\tau$, are determined by the
cross sections for WIMPs to annihilate and to scatter from
nuclei.  The equilibration timescale evaluates to
\begin{eqnarray}
\label{eqn:tau}
     \tau_\odot &=& 1.9\times10^5 \, {\rm yrs} \, [ \rho_{\chi,0.3}
     f(m_\chi)(\sigma_A v)_{26}]^{-1/2} \nonumber \\
     &\times& (m_\chi/100\,{\rm GeV})^{-3/4} \sigma_{40}^{-1/2}.
\end{eqnarray}
Here, $(\sigma_A v)_{26}$ is the annihilation cross section
(times relative velocity $v$ in the limit $v \rightarrow 0$) in
units of $10^{-26} \, {\rm cm}^{-3}{\rm s}^{-1}$. The
equilibration timescale for the Earth is obtained by replacing
the prefactor of Eq.~(\ref{eqn:tau}) by $1.1 \times 10^{8}$~yr.
Using the canonical numbers we have adopted, the equilibration
timescales for the Sun and Earth are both small compared with
the age of the Solar System, but the equilibration timescales
may vary by several orders of magnitude over reasonable ranges
of the WIMP parameter space (and even more if more exotic
physics, like a Sommerfeld \cite{Hisano:2004ds} or
self-capture enhancement \cite{Zentner:2009is}, is
introduced).  To illustrate, we show in Fig.~\ref{fig:PP} the
equilibration timescales, for various DD and $\nu$ID rates, for
realistic supersymmetric dark-matter candidates (using DarkSUSY
\cite{Gondolo:2004sc}).

Suppose now the WIMP model parameters are determined, 
e.g., from the LHC and/or by theoretical
assumption/modeling.  Then the unknowns in
Eqs.~(\ref{eqn:direct}) and (\ref{eqn:neutrinos}) will be
the halo density  $\rho_\chi$ and the number $N$ of WIMPs in the
Sun (or Earth).  The measured DD rate will then provide the
local halo density $\rho_\chi$.  Measurement of the $\nuID$ rate
will then determine $(N/\Neq) = \tanh(t/\tau)$ (in both the Sun
and the Earth).

\begin{figure}[t]
\resizebox{!}{7cm}{\includegraphics{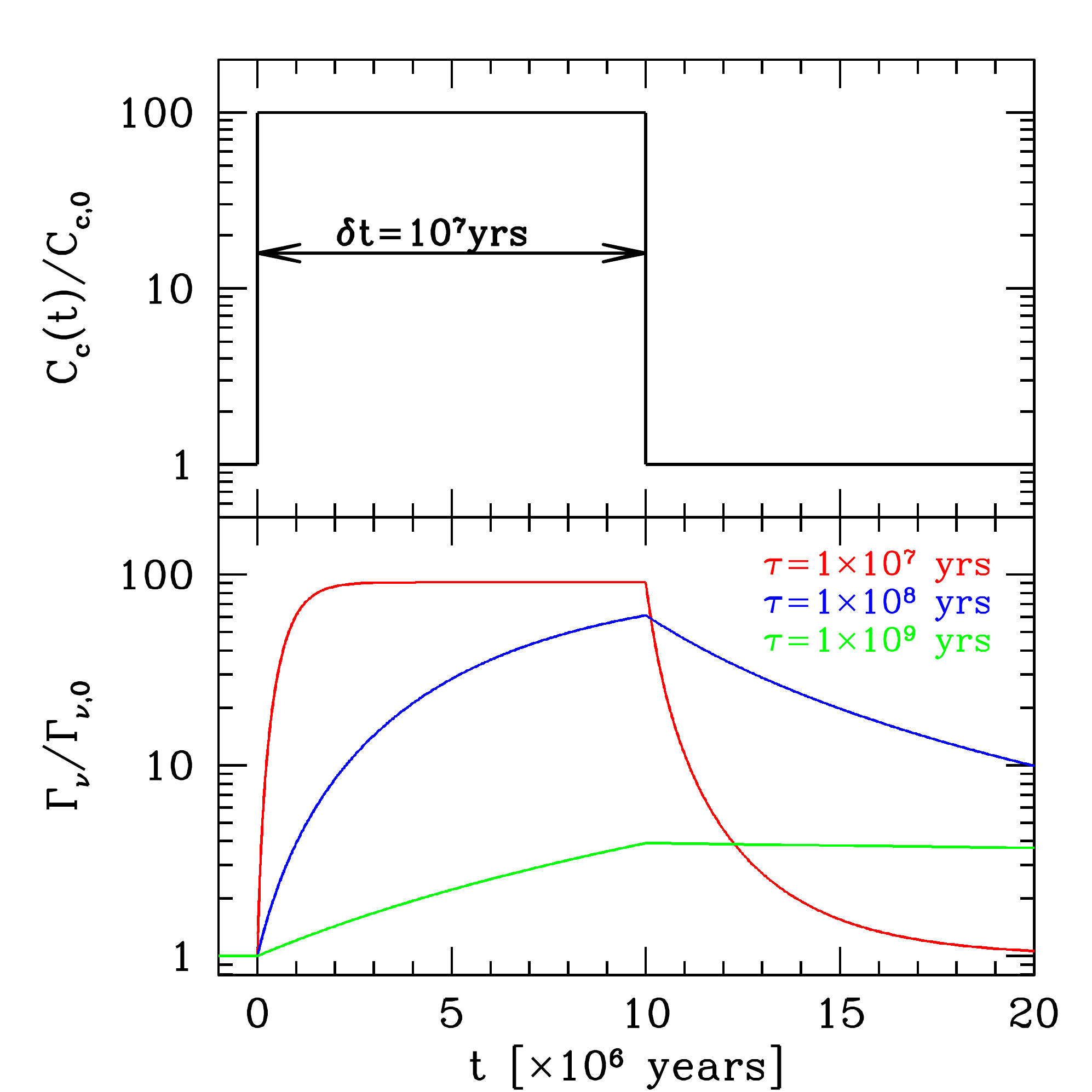}}
\caption {\small The neutrino-flux enhancement
     from an encounter, of duration $10^7$ yr, of the Solar
     system with a region where the dark-matter density is
     enhanced by a factor of 100 (e.g. a $10^9\,M_\odot$
     subhalo). The top panel shows the
     capture rate (i.e., the DD rate).  The
     bottom panel shows the resulting energetic-neutrino flux
     from such an encounter for three equilibration timescales.} 
\label{fig:history1}
\end{figure}

For example, suppose the equilibration timescale is $\tau_\odot
\approx 10^7$ years in the Sun, and that the Solar System
entered a region of density $\rho= 100\, \rhosun$ (where
$\rhosun=0.3$~GeV~cm$^{-3}$ is the smoothed local halo density)
a time $\delta t \approx 10^6$ years ago, e.g., a
$10^9\,M_\odot$ halo (see
Fig.~\ref{fig:history1}). We would then see a boosted DD rate
and a boosted energetic-neutrino flux
from the Sun, but the energetic-neutrino flux from the Earth
would be correspondingly weaker, since the equilibration time in
the Earth is longer.  Now, suppose that the Solar System exited
this high-density region a million years ago. The DD rate would
be at the canonical value, but the energetic-neutrino
fluxes from the Sun/Earth would still be boosted.  Finally,
suppose that the Solar System exited the high-density region
$10^7$ years ago. In that case, the DD rate and
energetic-neutrino flux from the Sun would have the canonical
values, but the neutrino flux from the Earth would still be
boosted, as $\tau_\oplus > \tau_\odot$.  

\begin{figure}[t]
\resizebox{!}{7cm}{\includegraphics{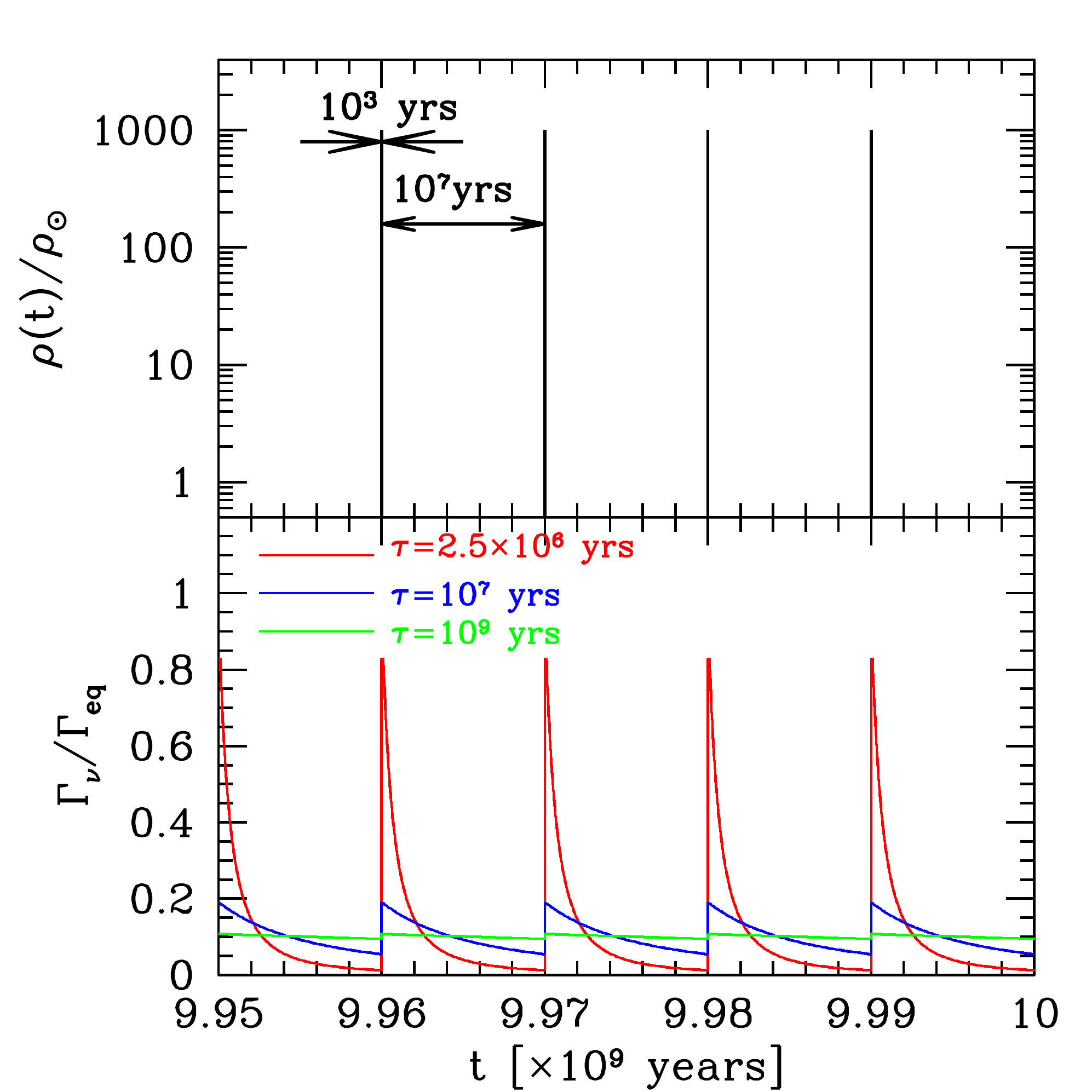}}
\caption {\small  The neutrino flux in a hypothetical scenario
     where all dark matter is in dense objects of $1
     M_\odot$. Different curves correspond to equilibration
     timescales as shown. Short
     equilibration timescales (e.g. Sun) almost deplete
     completely the amount of WIMPs in a time comparable to the
     timescale between interactions. Longer equilibration
     timescales (e.g. Earth) result in a constant elevated
     flux.}
\label{fig:history2}
\end{figure}

In reality, the capture rate $C_c(t)$ in Eq.~(\ref{eqn:riccati})
is a function of time, and the equation for the number of WIMPs
in the Sun or Earth can be integrated numerically to give the
annihilation rate $C_a N^2(t)/2$ as a function of
time.  To illustrate, imagine that all of halo dark matter was distributed
in objects of a single mass, $M=M_1\, M_\odot$, each with a
density $1000\,\beta\rhosun$ \cite{Koushiappas:2009du}. The
radius of these subhalos would then be $R = 0.1\,{\rm pc} \,
(M_{1}/\beta)^{1/3}$.  The transit time of the
Solar System through such an object is $\delta t \approx 1000
(M_{1}/\beta)^{1/3}\, {\rm yr}$.
The mean-free time between encounters with such objects is
$\bar{t} \approx 10^7\, (M_1 \beta^2)^{1/3}\, {\rm yr}$.
In this toy model, the dark-matter density (and
hence the DD rate) is zero unless the Solar System is within a subhalo.
Fig.~\ref{fig:history2} illustrates the effects from such a
scenario for $M_1=1$. For equilibration timescales which are of order $\tau
\sim \bar{t}$, the energetic-neutrino signal 
is depleted completely prior to the next encounter, while for
longer equilibration timescales, the net effect is an elevated
signal at all times. For example, if $\tau_\odot \sim \bar{t}$,
and $\tau_\oplus > \tau_\odot$, the signal from the Earth will
be boosted relative to the signal from the Sun for most of the
time.

\begin{figure}[t]
\resizebox{!}{4.2cm}{\includegraphics{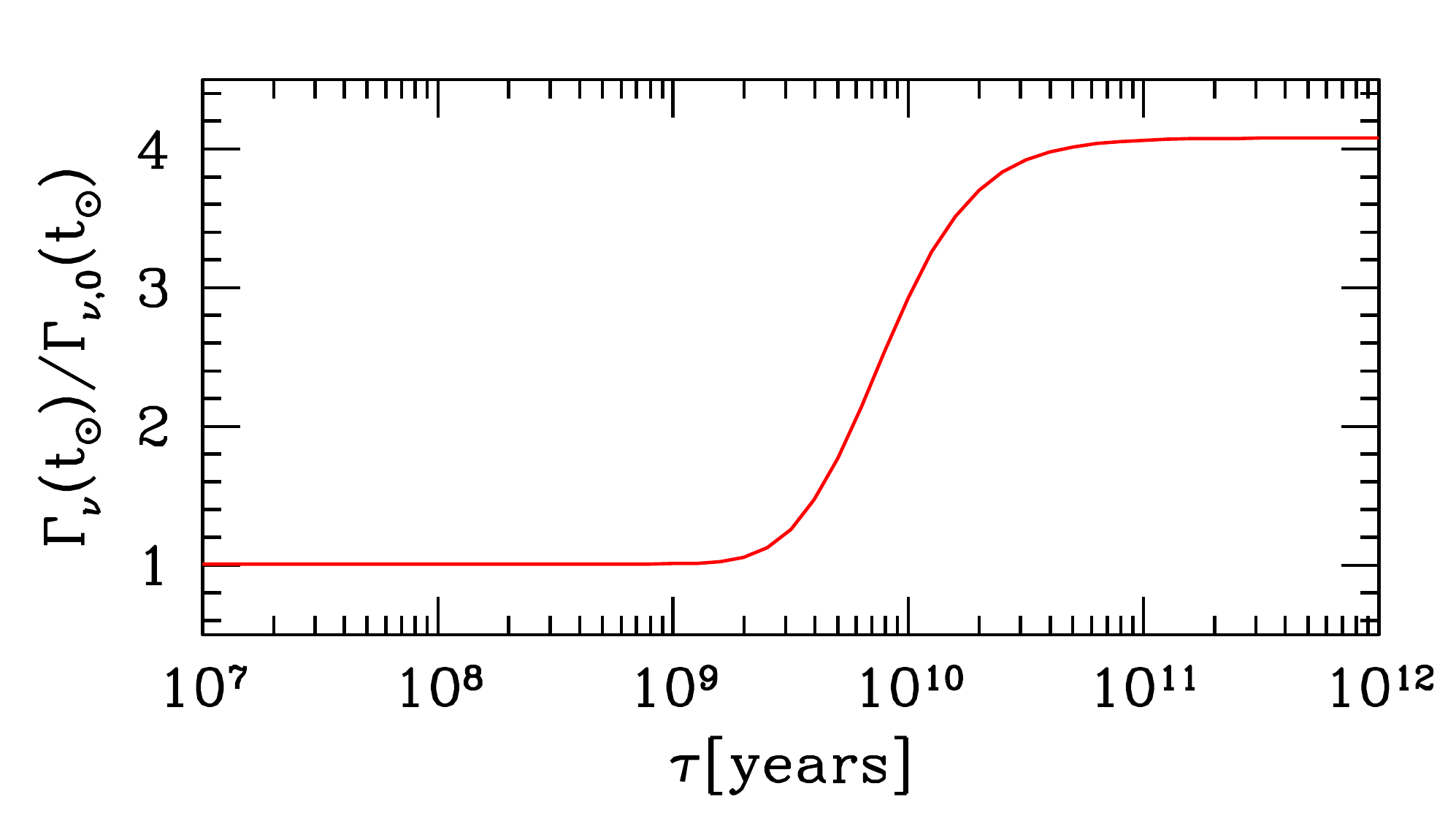}}
\caption {\small The net effect on the energetic-neutrino flux
     of the presence of $10^{-6} 
     M_\odot$ objects along the solar Galactic radius during the
     whole lifetime of the Solar System. Long equilibration
     timescales result in the net build-up of WIMPs and thus an
     increase in the neutrino flux relative to that which would
     be obtained in a smooth halo.} 
\label{fig:longtaus}
\end{figure}

Finally, substructure may also speed up the equilibration
between capture and annihilation in cases where the smooth-halo
equilibration timescale is larger than the age of the Solar
System.   Suppose the dark matter has a smooth component and
some substructure down to very small scales, $M\sim 10^{-6} \Msun$.  The
abundance of the smallest subhalos can be inferred by extrapolating
the subhalo mass function measured in simulations at larger mass
scales.  If we take the density within these $10^{-6}~M_\odot$
objects to be $\sim 100$ times the smooth 
value, the radius of these objects is $10^{-2}\, {\rm pc}$; the
crossing time is 50 years; and the mean time between encounters
is roughly a million years.  For equilibration timescales less
than the age of the Sun ($\tau_\odot \ll t_\odot$), the signal
will be roughly at the equilibrium value of the smooth component
for most of the time.  However, for long equilibration timescales
(e.g. the Earth), the amount of depletion between interactions
is negligible. This effect leads to a continuous build-up of
WIMPs in the Earth, augmented by brief periods of an increased
capture due to interactions with the subhalos. This results in
an energetic-neutrino signal that today is higher than the
signal that would be obtained from the smooth component.  This
can be understood as follows:  For $N \ll N_{\rm eq}$, the
second term in Eq.~(\ref{eqn:riccati}) is small.  While the
cross section for the Solar System trajectory to intersect
subhalos is $\propto \beta^{-2/3}$, the capture rate while in
them is $\propto \beta$, thus giving rise to a net increase in
the capture rate $\propto \beta^{1/3}$.  Fig.~\ref{fig:longtaus}
shows the net effect of this speed-up.

In summary, we considered the effects of Galactic substructure
on energetic neutrinos from WIMP annihilation in the Sun and the
Earth.  While DD experiments depend on the {\it
local} dark-matter density and velocity distribution, the
energetic-neutrino fluxes from the Sun and the Earth depend on
the past trajectory of the Solar System through the clumpy
Galactic halo.  If experimental DD and $\nuID$ signals are
obtained before the dark-matter particle-physics parameters are
known, then the potential for probing dark matter via the
DD/$\nuID$ ratios will be compromised by the particle-physics
uncertainties.  If, however, the particle-physics parameters
are known, then measurement of the DD rate and the
$\nuID$ rates from the Sun and Earth can be used to probe
Galactic substructure at the solar radius, with the
equilibration timescales setting roughly the mass scales that can
be probed.  

\smallskip
We acknowledge useful conversations with J.~Beacom,
I.~Dell'Antonio,  R.~Gaitskell,
A.~Geringer-Sameth, G.~Jungman, L.~Strigari, and A.~Zentner.
SMK thanks the Caltech/JPL W.~M.~Keck
Institute for Space Studies for hospitality during the preparation of
this article. The work of MK was supported by DoE
DE-FG03-92-ER40701 and the Gordon and Betty Moore Foundation.


\end{document}